\documentclass{iau}
\usepackage{graphicx}
\usepackage{inputenc}

\title[The Blazhko effect] 
{Blazhko effect in Cepheids \\ and RR Lyrae stars}

\author[R\'{o}bert Szab\'{o}]   
{R\'{o}bert Szab\'{o}}

\affiliation{Konkoly Observatory, Research Center for Astronomy and
Earth Sciences \\ of the Hungarian Academy of Sciences \\ Konkoly Thege Mikl\'{o}s \'{u}t 15-17. \\
H-1121, Budapest, Hungary \\email: {\tt szabo.robert@csfk.mta.hu}}

\pubyear{2013}
\volume{301}  
\pagerange{119--126}
\jname{Precision Asteroseismology, Celebration of the Scientific Opus of Wojtek Dziembowski}
\editors{W. Chaplin, J. A. Guzik, G. Handler \& A. Pigulski eds.}
\begin{document}

\maketitle

\begin{abstract}
The Blazhko effect is the conspicuous amplitude and phase 
modulation of the pulsation of RR Lyrae stars that was discovered 
in the early 20th century. The field of study of this mysterious 
modulation has recently been invigorated thanks to the 
space photometric missions providing long, 
uninterrupted, ultra-precise time series data.
In this paper I give a brief overview of the new observational 
findings related to the Blazhko effect, like extreme modulations, irregular 
modulation cycles and additional periodicities. I argue that these 
findings together with dedicated ground-based efforts now provide us 
with a fairly complete picture and a good starting point to theoretical 
investigations. Indeed, new, unpredicted dynamical phenomena have been discovered 
in Blazhko RR Lyrae stars, such as period doubling, high-order resonances, 
three-mode pulsation and low-dimensional chaos. These led to the proposal of a new 
explanation to this century-old enigma, namely a high-order resonance between 
radial modes. Along these lines I present the latest efforts and advances from the theoretical 
point of view. Lastly, amplitude variations in Cepheids are discussed. 
\end{abstract}

\firstsection 
\section{Introduction}

Although representatives of several pulsating variable types show amplitude modulation, 
the Blazhko effect of RR\,Lyrae stars is unique in many respects. First, the 
percentage of modulated fundamental mode RR\,Lyrae stars can be as high as 50\% 
(Konkoly Blazhko Survey, \cite[Jurcsik et al. 2009]{Jurcsik09}). 
Second, besides amplitude modulation, simultaneous phase modulation (or, equivalently period variation) 
is always present in modulated RRab stars (\cite [Benk\H o et al. 2010]{Benko10}). 
The third point is a freshly distilled lesson from a series of recent 
discoveries: the modulation of the high-amplitude nonlinear pulsational modes are accompanied 
by several dynamical phenomena (period doubling, resonances, chaos) and this feature
makes the Blazhko-effect in RR\,Lyrae stars a unique feature among the classes of pulsating variables. 
We note in passing that these dynamical effects have a special role in RR\,Lyrae stars, even the Blazhko 
effect itself might have a dynamical origin (ie. resonance between radial modes, see later). 

The Blazhko effect was discovered by S. Bla{\v z}ko more than a century ago 
(\cite[Bla{\v z}ko, 1907]{Blazhko07}) when he could not fit the period of RW\,Dra with a constant value  
(phase modulation). A few years later H. Shapley (\cite[Shapley, 1916]{Shapley16}) noticed the different heights 
of observed maxima of RR\,Lyrae, the prototype, discovering the amplitude modulation. The modulation time scale 
ranges from a few days to several years.

Only a few years have passed since the latest major review on the topic
(\cite[Kov\'{a}cs 2009]{Kovacs09}). By coincidence, the launch of {\it Kepler} 
(\cite[Borucki et al. 2010]{Borucki10}) preceded this excellent review only by a 
few months. This space telescope has opened a completely new window to the intricate behavior 
of the Blazhko-modulated RR\,Lyrae stars. Therefore I focus on the enormous progress that has 
become possible since \cite[Kov\'{a}cs (2009)]{Kovacs09}. 

A very important aspect is that the mathematics of the modulation is now well-understood, 
since without assuming any special physical mechanism, the mathematical description of the 
modulation of a carrier wave  (the pulsation in our case) can be elegantly derived from simple 
considerations (\cite[Benk\H o, Szab\'o \& Papar\'o 2011]{BSZP11}, \cite[Szeidl \& Jurcsik 2009]{SzJ09}). 
These descriptions admittedly do not convey any knowledge about the underlying physical mechanism, 
but help to discriminate between physical interpretations. In particular, instead of detecting only 
one side frequency (doublet) or symmetric side-peaks (triplet) (\cite[Alcock et al. 2003]{Alcock03}) or 
sometimes quintuplets (\cite[Hurta et al. 2008]{Hurta08}) around the dominant frequency and its 
harmonics from ground-based observations, with space data we can found a large (in theory infinite) number of 
side peaks in Blazhko RR\,Lyrae stars. For example in the Fourier-spectrum 
of V1127\,Aql up to 6-8 side peaks on both sides could be detected around the dominant frequency 
and its harmonics with CoRoT (\cite[Chadid et al. 2010]{Chadid10}). Thus, we are no longer limited by the rather 
high noise from ground-based observations. The side peak structure is only one aspect of the modulation. Many more 
unexpected results came from space photometric missions, and I continue with highlighting some of the spectacular 
{\it Kepler} observational results.

\section{The {\it Kepler} revolution - observational results}

The homogeneous photometric data set of more than 150,000 stars delivered by {\it Kepler} 
is unprecedented and proved to be a treasure trove both for transiting exoplanet finding and 
stellar astrophysics. The ultra-high precision (e.g. 8 ppm point per point precision for 
the bright, heavily saturated RR\,Lyr, the prototype, \cite[Kolenberg et al. 2011]{Kolenberg11}) 
exceeds by orders of magnitude what has been possible from the ground. In addition, the 
4-year-long, quasi-uninterrupted monitoring allowed the exploration of a completely new 
region of the parameter space. For many discoveries not the precision, but the continuity was the 
most important factor. The continuous observations were interrupted only for a few 
hours per month for data downloading or a few days in case of shorter and up to two weeks for longer 
technical problems. Some stars that fell on the dead Module 3 show regular, 3-month-long gaps 
annualy. 

With this extraordinary instrumental setup apart from a few RRc stars (Moskalik et al. 2013 in prep.) 
forty-four RRab stars were observed, 17 of them being Blazhko-modulated, which is close to 40\%.
A few more RR\,Lyrae stars have been discovered either serendipitiously (background contamination) or 
by meticulous investigations by dedicated citizen scientists (PlanetHunters) and will be added to the 
list soon. Unfortunately, after four years of operations Kepler has lost its second reaction wheel, which  
degrades the pointing stability, hence photometric precision, as well.

\subsection{Extreme modulations - modulation extremes}

One of the most intriguing feature that {\it Kepler} was able to show us is the huge variety of 
modulation shapes, and in some cases its cycle-to-cycle variations. There are extreme cases 
with large variability in the modulation cycles, while in other stars the modulation seems to be 
more or less regular. The poster child of the rapid variations in modulation characteristics
is V445\,Lyr (\cite[Guggenberger et al. 2012]{Guggenberger12}), where significant variations 
are observed from cycle to cycle. In addition, drastic light curve shape variation is evident:
in Blazhko-maxima the light curve is that of a normal RRab star, but in Blazhko-minima the 
amplitude is an order of magnitude less, and it transforms into a double-maxima, irregular 
light curve, that does not resemble to any RRab light variations. A sibling of this object 
is CoRoT\,105288363 (\cite[Guggenberger et al. 2011]{Guggenberger11}).

Another important revelation was the multiperiodic nature of the modulations. One of the best-observed  
example found by ground-based observations is CZ\,Lac (\cite[S\'{o}dor et al. 2011]{Sodor11}), which 
was followed by many more examples from {\it Kepler}. In fact based on the four-year long 
{\it Kepler} data it seems that most of the modulated RRab stars show multiple periodicities in their 
modulations, except the longest period ones, where more data would be necessary to establish any 
multi-periodicity (Benk\H o et al. in prep.). 
In CZ\,Lac two simultaneous modulation periods were detected and both changed its period from one observing 
season to the next. This fact along with the multiperiodic/irregular behavior of the modulation immediately 
raises the question whether the underlying dynamics is chaotic, and the changing, seemingly multiperiodic 
modulations are simply the manifestations of a chaotic behavior. There are indications that this scenario might 
indeed be plausible (see below).

\subsection{The unexpected beauty: period doubling}

The most unexpected and instrumental discovery on the Blazhko-front has been the detection of a 
well-known dynamical phenomenon, the period doubling (hereafter PD), which has never been observed 
in RR\,Lyrae stars before {\it Kepler}. PD was easily noticed in the first long-cadence (29.5-min) 
{\it Kepler} light curves (\cite[Kolenberg et al. 2010]{Kolenberg10}, \cite[Szab\'{o} et al. 2010]{Szabo10}) 
as the alternating maxima in some of the Blazhko RR\,Lyrae stars. PD manifests itself in the frequency domain 
as the presence of half-integer frequencies between the dominant pulsational frequency and its harmonics. 
Since the first detection, it turned out that the majority of Blazhko-modulated stars shows PD, at least 
temporarily. In each cases the strength of PD varies with time, and it can vanish for long time intervals, 
which partly explains why it could remain unnoticed before the space photometry era. Anyhow, the fact that PD 
is present in most of the Blazhko RRab stars, and has never been seen in non-modulated RR\,Lyrae stars even with 
the precision allowed by {\it Kepler} (\cite[Szab\'{o} et al. 2010]{Szabo10}, \cite[Nemec et al. 2011]{Nemec11}) 
demonstrates boldly that there should be a strong connection between PD and the Blazhko effect itself. 
In Fig.\,\ref{fig2} we see a characteristic and very strong PD phase of RR\,Lyrae, the eponym of its class.

\begin{figure}[t]
\begin{center}
 \includegraphics[angle=270,width=10cm,keepaspectratio=false]{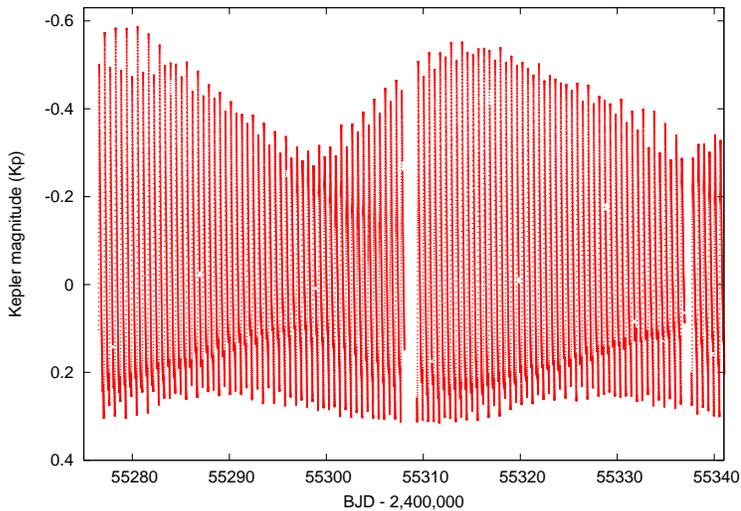} 
 \caption{Normalized Q5 short-cadence (1-min sampling) {\it Kepler} data of RR\,Lyrae, the prototype. Period doubling is 
persistent throughout this 2-month section. Notice the huge difference between the heights of consecutive maxima at the 
beginning. The plot contains more than 91,000 individual data points.}
   \label{fig1}
\end{center}
\end{figure}

\subsection{Additional periodicities in modulated RR Lyrae stars}

Interestingly, in addition to the traditional frequency solution of the Blazhko RR\,Lyrae stars (dominant mode, 
its harmonics and modulation side peaks), extra frequencies can be frequently found in Blazhko RR\,Lyrae stars 
if high enough precision space photometric data are available. These frequencies are present in non-Blazhko stars 
as well (\cite[Moskalik 2013]{Moskalik13} and Moskalik, these proceedings) and their origin (if pulsation) can be either radial 
(Benk\H o et al, these proceedings) or non-radial modes. However, in Blazhko RRab stars most of these frequencies 
tend to show up around the radial low-order overtones, namely around the predicted frequencies of the first (O1) and 
second (O2) overtones (see e.g. \cite[Benk\H o et al. 2010]{Benko10}). As it has been shown by Wojtek Dziembowski, nonradial 
modes can be present in RR\,Lyrae stars (\cite[Dziembowski 1977]{Dziembowski77}), and these modes are most easily excited 
in the vicinity of radial overtones (\cite[Van Hoolst et al. 1998]{vHoolst98}), so the radial/nonradial dilemma cannot be readily 
solved based on photometric data only. It is worth noting in this context that in at least one case 
(RR\,Lyrae itself) we were able to prove that the radial O1 mode is excited (see Sec.~\ref{triple}).

\section{The {\it Kepler} revolution - theoretical aspects}

How can these recent findings fit into a coherent theoretical framework? 
Will eventually all the pieces of the Blazhko jig-saw puzzle fall into place?
In what follows I present the latest theoretical advancements inspired by the 
discoveries discussed above.

\subsection{Resonances, resonances, resonances}

Resonances play a crucial role in Cepheids, but because of the narrower region in 
the parameter space occupied by RR\,Lyrae stars, resonances were thought to play a negligible 
role in these high-amplitude horizontal branch pulsators. Surprisingly 
\cite[Szab\'{o} et al. (2010)]{Szabo10} and \cite[Koll\'{a}th, Moln\'{a}r \& Szab\'{o} (2011)]{Kollath11}
were able to prove by computing hydrodynamic models that the origin of the PD is undoubtedly 
a high-order (9:2) resonance between the fundamental mode and a high-order {\it radial} 
(strange) overtone. This resonance is at work for a relatively large portion of the mass - luminosity -
effective temperature - metallicity parameter space. Although the $8^{\rm th}$ and the $10^{\rm th}$ 
radial overtones can also be trapped in the outer regions, hence not so heavily damped or even excited, 
a large survey of models demonstrated unambiguously that the ninth overtone is coupled very strongly to 
the fundamental mode and causes the period doubling. A series of period-doubling bifurcations can lead to chaos. 
We were able to show such bifurcation cascades in RR\,Lyrae models. Further full hydro computations showed that our 
models often approach other high-order resonant states, like 14:19, 20:27, etc., between the fundamental and the first 
overtone modes, creating a huge variety of complex dynamical behaviors. 

\subsection{Triple-mode states}\label{triple}

One such dynamical state is a three-mode condition. \cite[Moln\'{a}r et al. (2012)]{Molnar12} 
investigated the hydro models with PD that showed period-6 behavior as well, indicating 
that the model is temporarily can be close to the 3:4 resonance between the fundamental and 
first overtone. This period-6 characteristic was found in the {\it Kepler} observations of 
RR\,Lyr, the prototype. Based on the models the presence of the first overtone was predicted 
with low amplitude, and indeed Q5-Q6 {\it Kepler }data of RR\,Lyrae showed the O1 frequency 
with high significance (\cite[Moln\'{a}r et al. 2012]{Molnar12}). Based on this result at least 
some of the periodicities (especially around O1) should be the first radial overtone itself, 
but the presence of additional, non-radial modes cannot be completely ruled out. This three-mode 
state (fundamental, O1 and O9) on the one hand represents a new pulsational behavior dissimilar to 
the well-known double-mode (RRd) pulsators, and on the other hand enables an even more diverse 
bonanza of complex dynamical states, including chaos (\cite[Plachy et al. 2013]{Plachy13}). 

\subsection{Radial resonance as an explanation of the Blazhko effect}

\cite[Buchler \& Koll\'{a}th (2011)]{BK11} moved forward and showed using the amplitude 
equation formalism that if the 9:2 resonance between the fundamental mode and the $9^{\rm th}$ 
overtone is present, then in a large part of the parameter space regular, irregular, even chaotic 
modulations occur naturally. This result is a very important step towards the understanding of 
the Blazhko effect. The final step of proving the concept with full hydrodynamics models is 
still remains to be done, but it is worth mentioning here that \cite[Smolec \& Moskalik (2012)]{SM12} 
were able to produce modulated hydrodynamic models of BL\,Her stars in the presence of 
period doubling. Although much work is required to fully validate the resonance model, e.g. to 
reconcile the models with occurrence rates and observed quantities, working out 
Blazhko RRc models just to name a few, the radial resonance paradigm is currently the most tenable, 
and the only one that is backed up by full hydrodynamic models. Additional resonances and the 
presence of nonradial modes are also highly probable, and finding them should be of prime priority. 

\subsection{Low-dimensional chaos in the Blazhko modulation}

Based on a recent work of Plachy et al. (in prep.), there are indications that the 
modulation of some of the {\it Kepler} RRab stars with shorter modulations periods may be the result 
of low-dimensional chaos. The work is based on advanced mathematical methods, like global 
flow reconstruction and compares various return maps of the observed data to synthetic, chaos-generated 
data sets. The key factor for the successful application of the method is the observed number of modulation 
cycles. RR\,Lyrae stars having the shortest period (20-30 days) modulations accumulated enough 
cycles during {\it Kepler's} 
4-year operation to be suitable for such kind of analyses. For the long-period Blazhko stars more observational 
data would be essential. If {\it Kepler's} operation is extended and it continues to observe the same field in 
the two-wheel mode (e.g. to detect transit-timing variation of the already discovered planetary systems), then 
there is a good chance that a few {\it Kepler} Blazhko stars can be monitored as well (\cite[Moln\'{a}r 2013a]{Molnar13a}). 
We emphasize here again that the radial resonance model (\cite[Buchler \& Koll\'{a}th 2011]{BK11}) is able to predict 
chaotic modulation cycles, hence observations seem to confirm the theoretical predictions. 

\section{Amplitude variations in Cepheids}

Although several flavors of amplitude variation have been detected in Cepheids, 
these do not always resemble the Blazhko effect seen in RR Lyrae stars. 
One such example is the large number (19\%) of first overtone - second overtone 
(O1/O2) double-mode Cepheids in the Large Magellanic Cloud that show long-period 
($P_{mod} >$ 700d), anti-correlated amplitude modulation discovered by 
\cite[Moskalik \& Ko{\l}aczkowski (2009)]{MK09}. The true occurrence rate can be  
higher, considering the finite length of the OGLE-II observations and a significant 
number of O1/O2 double-mode Cepheids showing amplitude variations on even longer time 
scales. The origin of this phenomenon is currently unknown. 

Another prominent example is the amplitude change seen in the famous Cepheid, $\alpha$\,UMi 
or Polaris (\cite[Turner et al. 2005]{Turner05}). The light as well as the radial velocity 
amplitude of this overtone Cepheid had been decreasing until the mid-nineties then it rebounded 
and it has been increasing since then (\cite[Bruntt et al. 2008]{Bruntt08}, \cite[Spreckley \& 
Stevens 2008]{SS08}). It is not clear whether the variation is secular or cyclic. Obviously, 
more observations are required to answer this question. The cause of the 
variation is also a matter of debate, the suggested mechanisms range from variations in the 
stellar structure due to evolution through the presence of an additional (presumably) nonradial mode. 
Unless it is an extremely long-period modulation, there is no indication that the amplitude variation 
of Polaris would be related to the Blazhko-effect in RR Lyrae stars.

\begin{figure}[t]
\begin{center}
 \includegraphics[width=12.5cm]{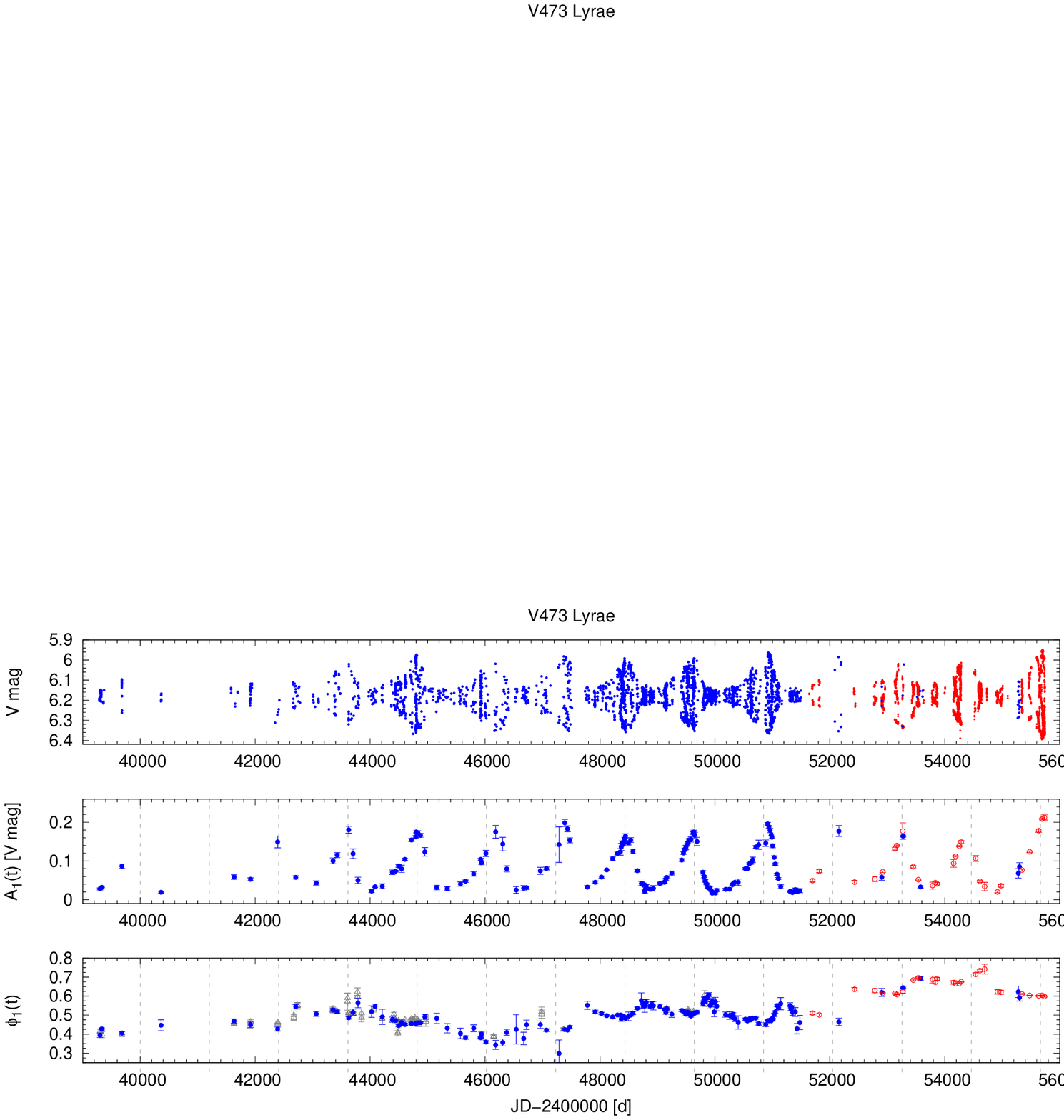} 
 \caption{Blazhko-like variations of V473\,Lyr. Upper panel: light curve, middle panel: amplitude variation 
($A_1$ term of the Fourier-decomposition), bottom panel: phase variation. The vertical dashed
lines denote the 1205-day modulation cycles. The full circles are V data, open symbols are scaled Str\"{o}mgren 
v observations to match the amplitudes in V. In the bottom plot the triangles are O-C data taken from 
the literature (\cite[Moln\'{a}r et al. 2013b]{Molnar13b} and references therein.)}
   \label{fig2}
\end{center}
\end{figure}

Contrary to Polaris, V473 Lyrae (\cite[Burki et al. 1986]{Burki86}) is a possible example for 
a Blazhko Cepheid. The characteristics of the light variation of this presumably second overtone 
Cepheid are almost identical to the Blazhko-modulated RRab stars (Moln\'{a}r et al. in prep.). The argument 
is based on a recent analysis of new photometric data (\cite[Moln\'{a}r et al. 2013b]{Molnar13b}) showing 
Blazhko-like amplitude modulation and simultaneous period variation, a distinctive characteristic of Blazhko 
RR Lyrae stars (see Fig.\,\ref{fig2}).

As a side note we add that the variations we discussed in this section fit the trend that was found on shorter 
time scales (Evans et al., these proceedings and references therein), namely the greater tendency for instability 
in the pulsation of  overtone pulsators compared to Cepheids pulsating in the fundamental mode.

\section{The future of the Blazhko research field}

From the observational side the success story of space photometry is expected to be continued. 
NASA's TESS (Transiting Exoplanet Survey Satellite, \cite[Ricker et al. 2010]{Ricker10}) mission 
with a launch scheduled for 2017 will monitor the whole sky (it will observe a given field for 27 days) 
searching for short-period planets around bright stars. The design of four large field-of-view camera 
will allow good photometric precision. A small fraction of the sky around the ecliptic poles will be 
observed continuously, that will be an excellent hunting ground for longer period Blazhko stars. 
The European counterpart of TESS is the PLATO (PLAnetary Transits and Oscillations of Stars) mission (\cite[Reuer et 
al. 2013]{Reuer13}). If selected, the arrangement of 32 normal and 2 fast telescopes on a 
common platform and the judiciously designed overlap in the monitored fields between the telescopes will 
provide large field-of-views and high precision. During the nominal 5-yr mission half of the sky could be 
covered with long (from months to up to a year) staring phases peppered with shorter step-and-stare intervals. 
The earliest launch date for PLATO is 2024. In both cases the missions will most probably avoid the overcrowded low 
galactic latitude regions. Needless to say, both space photometric projects can continue the pioneering work 
of CoRoT and {\it Kepler} and contribute to the understanding of the long-standing Blazhko-problem.

There is a great potential in spectroscopic observations, as well. Systematic, dedicated studies covering several 
Blazhko cycles are rare. Promising results have already been published recently, e.g. the detection of He\,I and He\,II 
lines in Blazhko stars (\cite[Preston 2009, 2011]{Preston09, Preston11}, 
\cite[Gillet et al. 2013]{Gillet13}, and also see the contributions of Guggenberger and Kolenberg, these proceedings).
Understanding the complicated dynamics of the pulsating atmosphere and the study of any non-spherical asymmetries (presence of nonradial 
modes for instance) would greatly benefit from such investigations. 

On the theoretical front, the one-dimensional hydrocodes, e.g. the Budapest-Florida code (\cite[Koll\'{a}th et al. 2002]{Kollath02}), 
or the Warsaw code (\cite[Smolec \& Moskalik 2008]{SM08}) still have a role in understanding physical concepts 
and mechanisms, especially in the light of the radial resonance paradigm. It has become increasingly clear that the 
multi-dimensional codes (Kupka, these proceedings, \cite[Geroux \& Deupree 2011, 2013]{Geroux11, Geroux13} and  
these proceedings) will greatly advance our understanding of stellar pulsation in general and the Blazhko effect in particular 
in the near future. When long simulations with many different initial conditions become feasible in two or three-dimensions, 
the self-consistent treatment of convection, the ability of modeling nonradial motions (or even modes) will make these 
hydrocodes excellent tools in helping to convey a better picture of the mode selection mechanism, the interaction between 
pulsation and convection, and the modulation mechanism itself. I strongly believe that these inherently three-dimensional objects 
and dynamical processes deserve the enormous work that is required to model them in multi-dimensions.

\begin{acknowledgments}
The author gratefully acknowledges the Lend\"ulet-2009 Young Researchers' Program and the J\'{a}nos Bolyai Research Scholarship 
of the Hungarian Academy of Sciences, the HUMAN MB08C 81013 grant of the MAG Zrt., the Hungarian OTKA grant K83790, 
the KTIA URKUT\_10-1-2011-0019 grant, the European Community's FP7/2007-2013 programme 
under grant agreement no. 269194 (IRSES/ASK), and the IAU for the travel grant. The inspiring 
discussions and influential work of Wojtek Dziembowski in the field of stellar pulsation are also thankfully acknowledged. 
Fruitful discussions with Z. Koll\'{a}th, L. Moln\'{a}r, E. Plachy, J. M. Benk\H o, K. Kolenberg, P. Moskalik, and R. Smolec are 
appreciated. The author wishes to dedicate this review to the memory of J. Robert Buchler.
\end{acknowledgments}

\end{document}